\documentclass[aps,twocolumn,pra]{revtex4}
\usepackage[dvips]{graphicx}
\def\be{\begin{equation}}
\def\ee{\end{equation}}
\def\bea{\begin{eqnarray}}          
\def\eea{\end{eqnarray}}
\def\bi{\begin{itemize}}
\def\ei{\end{itemize}}
\begin{document}
\title{$N$-particle Bogoliubov vacuum state}
\author{Jacek Dziarmaga and Krzysztof Sacha}
\address{Intytut Fizyki Uniwersytetu Jagiello\'nskiego, ul.~Reymonta 4, 
30-059 Krak\'ow, Poland}
\begin{abstract}

We consider the Bogoliubov vacuum state in the number-conserving Bogoliubov 
theory proposed by Castin and Dum [Phys. Rev. A {\bf 57}, 3008 (1998)]. 
We show that in the particle representation the vacuum 
can be written in a simple diagonal form. The vacuum state can describe
the stationary $N$-particle ground state of a condensate in a trap, but 
it can also represent a dynamical state when, for example, a Bose-Einstein 
condensate initially prepared in the stationary ground state is subject to 
a time-dependent perturbation. In both cases the diagonal form of the Bogoliubov 
vacuum can be obtained by basically diagonalizing the reduced single particle 
density matrix of the vacuum. We compare $N$-body states obtained within the 
Bogoliubov theory with the exact ground states in a $3$-site Bose-Hubbard
model. In this example, the Bogoliubov theory fails to accurately describe
the stationary ground state in the limit when $N\to\infty$ but a small fraction
of depleted particles is kept constant. 
\end{abstract}
\maketitle
\section{Introduction}
Bose-Einstein condensate (BEC) is a state of $N$ bosons where all  
particles occupy the same single particle state \cite{dalfovo99}. 
Such a state can be an eigenstate of the system only in an ideal case 
when particles do not interact. However,  dilute atomic gases offer 
a possibility for experimental realization of states that are not too far 
from ideal BEC \cite{carnellketterle}. In a dilute system the effect 
of interactions is weak and resulting quantum depletion of a condensate 
is not greater than 1\% in a typical experiment \cite{castinLesHouches}. 

From the point of view of theoretical description of a system 
weak interactions allow one to employ various perturbation approaches. The
approach commonly used in the field is the Bogoliubov theory which basically
relies on substitution of a full quantum many-body Hamiltonian by an effective 
Hamiltonian which is a quadratic approximation of the exact Hamiltonian
\cite{pines,dalfovo99}. 
In the original version  of  the 
Bogoliubov theory a condensate state is
identified with a coherent state that for a system 
of massive particles, in non-relativistic quantum mechanics, can not be 
acceptable. In the literature one may find several versions of number conserving 
Bogoliubov theory 
\cite{girardeau59,girardeau98,gardiner97,castin98,castinLesHouches}. 
In the present publication we consider 
the theory proposed by Castin and Dum \cite{castin98,castinLesHouches}.
In this introductory part we will briefly describe the key elements of the theory. 

Let us begin with a case where $N$ particles, interacting via
zero range potential $V(\vec r)=g_0\delta(\vec r)$, are trapped in a
time-independent potential $U(\vec r)$. We would like to get
approximation for the ground state of the system. 
In the Castin and Dum theory 
bosonic field operator $\hat\psi$ is decomposed 
into an operator $\hat a_0$ that annihilates a particle in a condensate mode 
$\phi_0$ and an annihilation operator $\delta\hat\psi$ defined in the subspace 
orthogonal to $\phi_0$,
\be
\hat\psi(\vec r)=\phi_0(\vec r)\hat a_0+\delta\hat\psi(\vec r).
\label{expd}
\ee
We are interested in $N$-particle states where a contribution 
coming from the subspace orthogonal to $\phi_0$ is small,
\bea
\langle \hat a_0^\dagger\hat a_0 \rangle & \approx & N, 
\cr
\int {\rm d}^3r\langle \delta\hat\psi(\vec r)^\dagger\delta\hat\psi(\vec r) 
\rangle & = & {\rm d}N,
\label{a0} 
\eea
where ${\rm d}N/N\ll 1$. Substituting (\ref{expd}) into the Hamiltonian of a
system and performing expansion in a "small parameter $\delta\hat\psi$" up to 
second order leads to the stationary Gross-Pitaevskii equation for $\phi_0$
\be 
H_{\rm GP}\phi_0=\left[-\frac{\hbar^2}{2m}\nabla^2+U(\vec r)
+g_0N|\phi_0|^2-\mu\right]\phi_0=0,
\ee 
and to an effective Hamiltonian quadratic in $\delta\hat\psi$. 
Castin and Dum have collected 
operators $\delta\hat\psi$ and $\hat a_0$ in such a way that the effective 
Hamiltonian, 
\be
\hat H_{\rm eff}=E_0(N)+\frac12 \int {\rm d}^3r \;(\hat\Lambda^\dagger,-
\hat\Lambda)
\;{\cal L}\;
\left(\begin{array}{c} \hat\Lambda \\ 
\hat\Lambda^\dagger\end{array}\right),
\label{heff}
\ee
depends only on operators 
\be
\hat\Lambda(\vec r)=\frac{\hat a_0^\dagger}{\sqrt{N}}\delta\psi(\vec r).
\ee
To get (\ref{heff}), sometimes it was necessary to
introduce by hand 
the operator $\hat a_0/\sqrt{N}$, i.e. a quantity of the order of one, see 
(\ref{a0}). Diagonalization of (\ref{heff}) reduces itself to diagonalization 
of a non-hermitian operator 
\be
{\cal L} = \left(
\begin{array}{cc}
H_{GP}+g_0N\hat Q|\phi_0|^2\hat Q   &   
g_0 N \hat Q\phi_0^2\hat Q^*        \\
-g_0 N \hat Q^*\phi_0^{*2}\hat Q    &   
-H_{GP}^*-g_0N \hat Q^*|\phi_0|^2\hat Q^*
\end{array}
\right),
\label{l}
\ee
where $\hat Q=1-|\phi_0\rangle\langle\phi_0|$.
Generally the ${\cal L}$ operator is diagonalizable and, for the ground state of 
the
system we are interested in, its eigenvalues $E_m$ are real. 
Owing
to the symmetries of ${\cal L}$ (i.e. $\sigma_x{\cal L}\sigma_x=-{\cal L}^*$ 
and 
$\sigma_z{\cal L}\sigma_z={\cal L}^\dagger$, where $\sigma_x$, $\sigma_z$ are
Pauli matrices) it is not difficult to show that if 
\be
|\psi_m^{\rm R}\rangle=\left(\begin{array}{c} 
u_m \\ 
v_m
\end{array}\right),
\ee
is a right eigenvector of ${\cal L}$ corresponding to an eigenvalue $E_m$, 
the left eigenvector is
\be
|\psi_m^{\rm L}\rangle=\sigma_z |\psi_m^{\rm R}\rangle,
\ee
and 
\be
\sigma_x |\psi_m^{\rm R}\rangle^*,
\label{fam2}
\ee
is a right eigenvector corresponding to $-E_m$. These properties
imply that the eigenvectors can be divided into two classes. The first class
(the so-called family "+") consists of eigenvectors which 
can be normalized so that 
\be
\langle\psi_n^{\rm R}|\sigma_z|\psi_m^{\rm R}\rangle=
\langle u_n|u_m\rangle-\langle v_n|v_m\rangle=\delta_{nm}.
\label{fpiden} 
\ee
Employing transformation (\ref{fam2}) we obtain, from 
eigenvectors belonging to the family "+", eigenvectors that form the family "$-
$" 
where $\langle\psi_i^{\rm R}|\sigma_z|\psi_j^{\rm R}\rangle=-\delta_{ij}$.
There are always two eigenvectors of ${\cal L}$
\bea
\left(\begin{array}{c} 
\phi_0 \\ 
0
\end{array}\right), &\;&
\left(\begin{array}{c} 
0 \\ 
\phi_0^*
\end{array}\right),
\eea
corresponding to zero eigenvalue. Collecting all eigenvectors of $\cal L$ one 
can decompose the identity operator in the following form
\bea
\hat 1&=&\left(\begin{array}{c} |\phi_0\rangle \\ 0 
\end{array}\right)(\langle\phi_0|,0)
+\left(\begin{array}{c} 0 \\ |\phi_0^*\rangle 
\end{array}\right)(0,\langle\phi_0^*|)
\cr &&
+\sum_{m\in"+"}\left[\left(\begin{array}{c} |u_m\rangle \\ |v_m\rangle 
\end{array}\right)
(\langle u_m|,\langle-v_m|)\right. \cr
&&\left.+
\left(\begin{array}{c} |v_m^*\rangle \\ |u_m^*\rangle \end{array}\right)
(\langle -v_m^*|,\langle u_m^*|)\right].
\eea

From the analysis performed it is clear that
\bea
\langle u_m|\phi_0 \rangle = 0, \cr
\langle v_m|\phi_0^{*} \rangle = 0,
\label{orth}
\eea
and in the subspace of functions orthogonal to a condensate wave function 
$\phi_0$ the completeness relation can be written in the following form
\be
\hat 1=
\sum_{m\in"+"}\left(|u_m\rangle \langle u_m|-
 |v_m^*\rangle\langle v_m^*|\right).
\label{complet}
\ee

Having calculated Bogoliubov modes $(u_m,v_m)$ one can easily find diagonal 
form of the effective Hamiltonian
\be
\hat H_{\rm eff}=\tilde E_0(N) + \sum_{m \in"+"}E_m\;\hat
b_m^\dagger\;\hat b_m,
\label{heffd}
\ee
where so called quasi-particle annihilation operators
\be
\hat b_m=\frac{1}{\sqrt{N}}\left(\hat a_0^\dagger\langle 
u_m|\hat\delta\psi\rangle
-\hat a_0\langle v_m|\hat\delta\psi^\dagger\rangle\right),
\label{bop}
\ee
fulfill a bosonic commutation relation
\be
[\hat b_m, \hat b_m^\dagger]=\delta_{nm}+{\cal O}(1/N).
\label{comb}
\ee
The ground state of the Hamiltonian (\ref{heffd}) is a Bogoliubov vacuum state 
that is annihilated by all quasi-particle annihilation operators
\be
\hat b_m|0\rangle_{\rm B}=0.
\label{bvac}
\ee
Excited states can be generated by means of the 
quasi-particle creation operators
\be
|m_1,m_2,...\rangle=\prod_{k=1}^\infty 
\frac{\left(\hat b_k^\dagger\right)^{m_k}}{\sqrt{m_k !}}|0\rangle_{\rm B}.
\label{focksp}
\ee

The advantage of the number conserving version of the Bogoliubov theory with
respect to the original theory is that one always deals with states of exactly 
$N$ particles --- application of quasi-particle operators does not change number 
of particles thanks to the presence
of the $\hat a_0$ and $\hat a_0^\dagger$ operators in Eq.(\ref{bop}).

Having a condensate in the ground state of the trapping potential $U(\vec r)$ 
(i.e. in the Bogoliubov vacuum state $|0\rangle_{\rm B}$)
we may apply a time-dependent perturbation. Then $\phi_0$ starts evolving
according to the time-dependent Gross-Pitaevskii equation and the Bogoliubov 
modes
evolve according to the time-dependent Bogoliubov-de Gennes equations
\be
i\hbar\frac{{\rm d}}{{\rm d}t}
\left(\begin{array}{c} u_k(\vec r,t) \\ 
v_k(\vec r,t) \end{array}\right)={\cal L}(t)
\left(\begin{array}{c} u_k(\vec r,t) \\ 
v_k(\vec r,t) \end{array}\right).
\label{tbdg}
\ee
However, the system remains in a Bogoliubov vacuum state but this Bogoliubov 
vacuum and the quasi-particle operators
$\hat b_m(t)$ depend on time. 

Actually one may define any initial Bogoliubov vacuum state (not necessary the
solution for the ground state of a time-independent problem). 
To this end 
one has to define $\phi_0$ and initial Bogoliubov modes $(u_m,v_m)$ in such a 
way that 
(\ref{fpiden}) and (\ref{orth}) are fulfilled. 
For example choosing $\phi_0=\psi_0$ and 
$(u_m=\psi_m,v_m=0)$ for the family "+" 
(where $\psi_j$ are the eigenstates of the harmonic oscillator) the
initial Bogoliubov vacuum corresponds to a perfect condensate where all
particles occupy the ground state wavefunction of the harmonic oscillator.

We should mention that in practice the projection operator $\hat Q$ 
in the $\cal L$ operator (\ref{l}) can be omitted. 
One can solve the Bogoliubov-de Gennes
equations in the standard form and at the end apply the operator $\hat Q$
\cite{castin98,castinLesHouches}.

\section{Results}

In this section we will restrict ourselves to analysis of 
the Bogoliubov vacuum state. The Bogoliubov
vacuum is a state of $N$ particle system where there is no quasi-particle, 
i.e. it is annihilated 
by all quasi-particle annihilation operators $\hat b_m$ (quasi-particle vacuum), 
see (\ref{bvac}). 
The particle vacuum (a state with no particles) is also annihilated 
by all $\hat b_m$ because of the presence of a particle annihilation 
operator in 
each part of the $\hat b_m$ operator, see (\ref{bop}). 
We have shown \cite{dziarmaga03pra,dziarmaga03jpb,dsJPB05}
that the Bogoliubov vacuum can be obtained
from the particle vacuum with the help of some particle creation operator
$\hat d^\dagger$,
\be
|0\rangle_{\rm B}\sim \left(\hat d^\dagger\right)^M |0\rangle.
\ee
Indeed, if we require that the $\hat d^\dagger$ commutes with all 
quasi-particle operators,
\be
[\hat b_m,\hat d^\dagger]=0, 
\label{setofe}
\ee
then 
\be
\hat b_m\left(\hat d^\dagger\right)^M |0\rangle=\left(\hat d^\dagger\right)^M 
\hat
b_m|0\rangle=0.
\ee
It turns out that two-particle creation operator 
\be
\hat d^\dagger=\hat a_0^\dagger\hat a_0^\dagger +\sum_{\alpha,\beta=1}^\infty 
Z_{\alpha\beta}\hat a_\alpha^\dagger
\hat a_\beta^\dagger,
\label{dop}
\ee
(where $\hat a_\alpha^\dagger$'s create atoms in modes $\varphi_\alpha$'s 
orthogonal to the condensate wavefunction $\phi_0$) 
solves the set of equations (\ref{setofe}). Substituting (\ref{dop}) to 
(\ref{setofe}) yields the following equation for the  
$Z_{\alpha\beta}$ matrix
\be
\langle v_m|\varphi_\alpha^*\rangle=
\sum_{\beta=1}^\infty
\langle u_m| 
\varphi_{\beta}\rangle 
Z_{\beta\alpha}.
\label{zeq}
\ee

One may wonder how it is possible that a bosonic quasi-particle annihilation
operator may annihilate, in fact, an infinite number of states (for each $N$ 
there 
is a Bogoliubov vacuum) \cite{yukalov}. It is due to the fact that, strictly 
speaking, $\hat b_m$ are not bosonic operators. However, for large $N$ 
corrections
to bosonic commutation relations are small [see (\ref{comb})] and at the present
order of approximation we may consider $\hat b_m$ as bosonic operators that
allows us to define (within a subspace of a fixed $N$) 
a Fock space (\ref{focksp}).

Let us now show that solution of (\ref{zeq}) can be significantly simplified 
thanks to properties of the Bogoliubov modes
\cite{dsJPB05}. To this end let us multiply
(\ref{zeq}) by $
\langle\varphi_\gamma^*|v_m\rangle$ and then sum over $m$ which leads to the 
following equation
\be
\langle \varphi_\gamma^*|{\rm
d}\hat\rho|\varphi^*_{\alpha}\rangle=\sum_{\beta=1}^\infty
\langle \varphi_\gamma^*|\hat\Delta|\varphi_{\beta}\rangle 
Z_{\beta\alpha},
\label{zeq1}
\ee
where ${\rm d}\hat\rho$ and $\hat \Delta$ operators are defined as
\bea
{\rm d}\hat\rho&=&\sum_{m\in"+"}|v_m\rangle\langle v_m |,\cr
\hat \Delta&=&\sum_{m\in "+"}
|v_m\rangle\langle u_m|.
\eea
Single particle density matrix corresponding to the Bogoliubov vacuum 
state reads
\be
\rho(\vec r,\vec r\;') =
N_0~\phi_0^*(\vec r)\phi_0(\vec r~')+
\sum_{m\in"+"} v_m(\vec r) v_m^*(\vec r~'),
\ee
which indicates that ${\rm d}\hat\rho$ operator is a part of the single particle 
density operator  corresponding to the subspace orthogonal to the condensate 
wavefunction $\phi_0$. Thus eigenstates of the single particle density matrix
\be
\rho(\vec r,\vec r\;') =N_0~\phi_0^*(\vec r)\phi_0(\vec 
r~')+\sum_{\alpha=1}^\infty {\rm d}N_\alpha~
\phi_\alpha^*(\vec r)\phi_\alpha(\vec r~'),
\ee
(where ${\rm d}N_\alpha$ are numbers of particles depleted from a condensate)
diagonalize also ${\rm d}\hat\rho$
\be
{\rm d}\hat\rho=\sum_{\alpha=1}^\infty {\rm d}N_\alpha~
|\phi_\alpha^*\rangle\langle\phi_\alpha^*|.
\ee

Employing the completeness relation (\ref{complet}) and the orthogonality 
relation
(\ref{fpiden}) it is easy to show that 
\be
{\rm d}\hat\rho \hat\Delta = 
\hat\Delta{\rm d}\hat\rho^*, 
\ee
which in turn implies that eigenstates of the single particle density matrix 
allows
us to find also diagonal form of the $\hat\Delta$ operator
\be
\hat \Delta=\sum_{\alpha=1}^\infty
\Delta_\alpha
|\phi_\alpha^*\rangle\langle\phi_\alpha|.
\ee
There is another equality fulfilled by the 
${\rm d}\hat\rho$ and $\hat \Delta$ operators 
\be
\hat\Delta^*\hat\Delta={\rm d}\hat\rho^*(1+{\rm d}\hat\rho^*),
\ee
which allows us to get simple relation between absolute value of $\Delta_\alpha$
 and a number
of particles ${\rm d}N_\alpha$  depleted from a condensate:
\be
|\Delta_\alpha|^2={\rm d}N_\alpha(1+{\rm d}N_\alpha).
\ee

Let us turn back to the equation (\ref{zeq1}). If we choose as the basis vectors 
$\varphi_\alpha$  the eigenstates $\phi_\alpha$ 
of the single particle density matrix we will obtain ${\rm d}\hat\rho$ and 
$\hat \Delta$ in diagonal forms and consequently also $Z_{\alpha\beta}$ 
matrix in a diagonal form,
\be
Z_{\alpha\beta}=\lambda_\alpha\delta_{\alpha\beta}, 
\ee
where
\be
\lambda_\alpha=\frac{{\rm d}N_\alpha}{\Delta_\alpha}.
\ee
Note that 
\be
|\lambda_\alpha|=\sqrt{\frac{{\rm d}N_\alpha}{{\rm d}N_\alpha+1}}.
\ee
Finally the Bogoliubov vacuum state in the particle representation can be
written in a nice diagonal form
\be
|0\rangle_{\rm B}\sim 
\left(\hat a_0^\dagger\hat a_0^\dagger +\sum_{\alpha=1}^\infty 
\lambda_\alpha\hat a_\alpha^\dagger
\hat a_\alpha^\dagger\right)^{N/2}|0\rangle,
\ee
where $\hat a_\alpha^\dagger$ operators create particles in modes that are 
eigenstates of the 
single particle density matrix.

The analysis performed shows beautiful properties of the Bogoliubov modes. 
Almost all
we need in order to obtain the Bogoliubov vacuum is given by eigenstates and 
eigenvalues of
the single particle density matrix. The only things that can not be determined 
from
this matrix are the phases of $\Delta_\alpha$.

\section{Comparison of the Bogoliubov theory with exact solutions for a model
system}

Knowledge of the Bogoliubov vacuum state in the particle representation allows
for precise comparison of the Bogoliubov theory with exact
solutions for model systems. 
In Ref.\cite{dziarmaga03pra} we have analyzed $N$ particles in a double well 
potential
within the tight binding approximation where we have found that, quite surprisingly, 
the Bogoliubov theory gives extremely good predictions for the ground state 
and low lying excited states in the entire range of the system parameters. This 
is
rather an exception than a rule and therefore in order to test the
theory we will focus now on a slightly more complex system, i.e. $N$ particles 
in a triple well potential. Restricting the single particle Hilbert space to 
the ground states in each well only, the Hamiltonian of the system reads
\be
H=
-\Omega
\sum_{\langle i,j \rangle}
\hat c_i^{\dagger}\hat c_j+
\frac12\sum_{j=1}^3\hat c_i^{\dagger}\hat c_i( \hat c_i^{\dagger}\hat c_i-1),
\label{Hubbard3w}
\ee
where $\hat c_i$ denotes bosonic operator that annihilates a particle in
an $i$-well, $\Omega$ stands for tunneling rate between neighboring wells.
This is the only parameter because the particle interaction 
coefficient was chosen as energy scale.
For $\Omega\gg N$ the ground state of the system is a BEC where all atoms
occupy the same condensate wavefunction \cite{hubard}. On the other hand
$\Omega\ll 1/N$ defines the Mott insulator regime where the Bogoliubov 
theory should definitely stop working. 

\begin{figure}
\includegraphics[width=8.6cm]{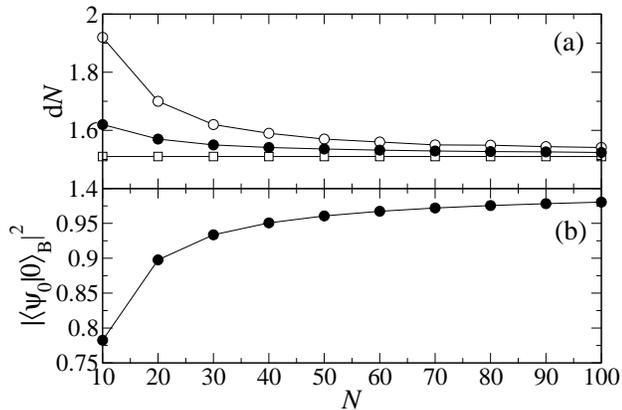}
\caption{
Panel (a) shows number of atoms depleted from a condensate wavefunction, 
i.e. ${\rm d}N$,
versus $N$ for $\Omega=0.01N$, calculated in the exact state (full circles), 
calculated directly in the Bogoliubov vacuum state in the particle
representation (open circles) and calculated with the help of Eq.~(\ref{dnbog})
(open squares). In panel (b) the squared overlap between the exact ground state
of the system $|\psi_0\rangle$ and the Bogoliubov vacuum $|0\rangle_{\rm B}$
in the particle representation is
presented.
}
\label{one}
\end{figure}

\begin{figure}
\includegraphics[width=8.6cm]{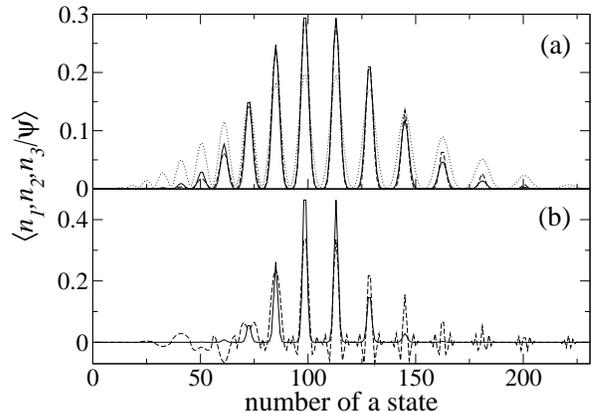}
\caption{
Panel (a): the solid line corresponds to the exact ground state $|\psi_0\rangle$, 
the dashed line to the Bogoliubov prediction $|0\rangle_{\rm B}$ and the dotted 
line to the perfect condensate state $|N:\phi_0\rangle$ where
all atoms occupy the condensate mode (\ref{phizero}), for $N=20$ and 
$\Omega=1$ (${\rm d}N=0.4$, $|\langle \psi_0|0\rangle_{\rm B}|^2=0.99$
and $|\langle \psi_0|N:\phi_0\rangle|^2=0.83$). 
Panel (b): the solid line corresponds to the exact ground state $|\psi_0\rangle$ and 
the dashed line to the Bogoliubov prediction $|0\rangle_{\rm B}$ for $N=20$ and 
$\Omega=0.1$ (${\rm d}N=2.5$, $|\langle \psi_0|0\rangle_{\rm B}|^2=0.76$).
The states are projected on $|n_1,n_2,n_3\rangle$ where $n_i$ is a number
of particles in an $i$-well.
}
\label{two}
\end{figure}

The ground state of the Gross-Pitaevskii equation corresponding to the
triple-well system reads
\be
\phi_0=\frac{1}{\sqrt{3}}\left(1,1,1\right)~
\label{phizero}
\ee
and the solutions of the Bogoliubov-de Gennes equations are 
\cite{dziarmaga03pra}
\bea
u_{\pm}=\frac{X\phi_{\pm}}{\sqrt{X^2-1}}~,~~v_{\pm}=\frac{-
\phi_{\pm}}{\sqrt{X^2-1}}~,
\nonumber\\
\eea
where
\bea
&&
\phi_+=
\frac{1}{\sqrt{3}}
\left( 1 , e^{i2\pi/3} , e^{-i2\pi/3} \right)~,
\nonumber\\
&&
\phi_-=
\frac{1}{\sqrt{3}}
\left( 1 , e^{-i2\pi/3} , e^{i2\pi/3} \right)~
\label{plusminus}
\eea
and
\be
X=\left(1+\frac{9\Omega}{N}\right)+
\sqrt{ \left(1+\frac{9\Omega}{N}\right)^2 - 1 }.
\ee
Diagonal form of the Bogoliubov vacuum in the particle representation is 
defined by the $\hat d^\dagger$ operator
\be
\hat d^\dagger= \hat a_0^\dagger \hat a_0^\dagger - 
\frac{1}{X} \hat a_1^\dagger \hat a_1^\dagger +
\frac{1}{X} \hat a_2^\dagger \hat a_2^\dagger,
\ee
where $\hat a_0^\dagger$ creates a particle in the condensate mode $\phi_0$
while $\hat a_1^\dagger$ and $\hat a_2^\dagger$ create particles in 
$\phi_1=(\phi_++\phi_-)/\sqrt{2}$ and $\phi_2=(\phi_+-\phi_-)/\sqrt{2}$
modes, respectively \cite{dziarmaga03pra}.
Number of atoms depleted from the condensate mode can be directly calculated 
from the Bogoliubov vacuum in the particle representation but 
it is given approximately by 
\bea
{\rm d}N&=&\int dx~\langle\delta\hat\psi^{\dagger}\delta\hat\psi\rangle \approx
\int dx~\langle\delta\hat\psi^{\dagger}\hat a_0
\frac{1}{N}\hat a_0^{\dagger}\delta\hat\psi \rangle\cr 
&=&
\int dx~\langle\hat\Lambda^{\dagger}\hat\Lambda\rangle
=\langle v_+|v_+\rangle+\langle v_-|v_-\rangle \cr
&=&\frac{2}{X^2-1}.
\label{dnbog}
\eea

\begin{figure}
\includegraphics[width=8.6cm]{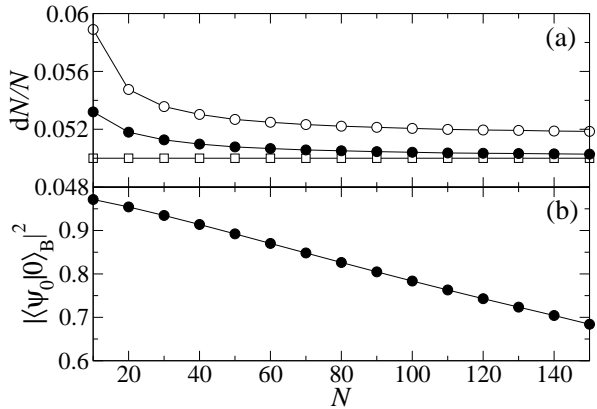}
\caption{
Panel (a) shows fraction of depleted atoms 
from a condensate wavefunction, i.e. ${\rm d}N/N$,
versus $N$ for $\Omega$ chosen so that ${\rm d}N/N\approx2/[N(X^2-1)]=0.05$, 
calculated in the exact state (full circles), 
calculated directly in the Bogoliubov vacuum state in the particle
representation (open circles) and calculated with the help of Eq.~(\ref{dnbog})
(open squares). In panel (b) the squared overlap between the exact ground state
of the system $|\psi_0\rangle$ and the Bogoliubov vacuum $|0\rangle_{\rm B}$
in the particle representation is presented.
}
\label{three}
\end{figure}

Increasing $N$ but keeping fixed number of atoms depleted from the
condensate mode one approaches the limit where the Bogoliubov theory should
become more and more accurate. We have done calculations for different $N$
but for fixed $\Omega/N$ chosen so that ${\rm d}N\approx 2/(X^2-1)=1.5$. 
The squared overlap between the Bogoliubov vacuum state and the exact ground 
state of the Hamiltonian (\ref{Hubbard3w}) is depicted in Fig.~\ref{one} where
one can indeed observe that with increasing $N$ the Bogoliubov solution
approaches the exact one. With increasing $N$ also number of depleted atoms 
calculated directly from the Bogoliubov vacuum state and with the help of
the estimation (\ref{dnbog}) coincide with the exact results.
Example of the Bogoliubov 
and exact states for $N=20$ an for two values of $\Omega$ are shown 
in Fig.~\ref{two}. When $\Omega$ is sufficiently large
we get perfect agreement between the Bogoliubov and exact results.

It is generally believed that when the fraction of atoms depleted from a
condensate is small, say a few percent, the Bogoliubov theory is reliable.
Let us now investigate the limit when $N$ increases but ${\rm d}N/N$ remains
constant. In Fig.~\ref{three} we present the results for different $N$
but for ${\rm d}N/N\approx 2/[N(X^2-1)]=0.05$. One can see that the greater $N$
the worse the Bogoliubov predictions. Actually increasing $N$ and 
keeping ${\rm d}N/N$ fixed we go far and far away from 
the BEC regime where $\Omega \gg N$. Indeed,
\be
\frac{{\rm d}N}{N} \rightarrow \frac{\sqrt{2}}{6} \frac{1}{\sqrt{\Omega N}}
+{\cal O}\left(1/N\right), 
\ee
and in order to keep ${\rm d}N/N$ constant the $\Omega$ must behave like $1/N$.
This example shows that even if the fraction of depleted 
atoms is very small and single particle density matrix can be estimated quite
satisfactory, for very large number of atoms the Bogoliubov theory fails
to predict structure of the $N$-body ground state. Basing on the triple-well 
problem only it is dificult to judge how general is such a behaviour.

\section{Conclusions}

We have considered the Bogoliubov vacuum states in 
the number-conserving version of the Bogoliubov theory. It is shown that 
the vacuum can be obtained in a simple diagonal form in the particle
representation. To this end one has to basically diagonalize single particle
density matrix of a system only. 
The Bogoliubov vacuum can describe eigenstates of a time
independent system and also a dynamical case where starting with a system in 
the ground state particles are perturbed by time dependent force. 

$N$-body states obtained with the help of the Bogoliubov theory have been 
confronted with exact solutions for a triple-well system. 
In the limit of $N\rightarrow\infty$ and ${\rm d}N=\rm const.$, 
the Bogoliubov vacuum states approach the exact solutions as
expected.
However, it turns out that 
in the limit of $N\rightarrow\infty$ and for arbitrary small but fixed 
${\rm d}N/N$, the Bogoliubov theory is not able to predict $N$-body states
of the system.

\section*{ Acknowledgments }

The work of JD was supported in part by Polish Government scientific 
funds (2005-2008) as a research project. KS was
supported by the KBN grant PBZ-MIN-008/P03/2003. 


\end{document}